\documentclass[11pt]{article}

 \newcommand{\be}{\begin{equation}}
 \newcommand{\ee}{\end{equation}}
 \newcommand{\ba}{\begin{eqnarray}}
 \newcommand{\ea}{\end{eqnarray}}
 \newcommand{\bl}{\begin{equation}\begin{array}{ll}}
 \newcommand{\el}{\end{array}\end{equation}}
 \newcommand{\bll}{\begin{equation}\begin{array}{lll}}
 \newcommand{\bdm}{\begin{displaymath}}
 \newcommand{\edm}{\end{displaymath}}
 \def\bea{\begin{eqnarray}}
 \def\eea{\end{eqnarray}}
 \def\barr{\begin{array}}
 \def\earr{\end{array}}
\def\p{\partial}
\def\d{\partial}
\def\dif{\partial}

\def\f{\varphi}
\def\ve{\varepsilon}
\def\ep{\epsilon}
 \def\De{\Delta}

\def\half{\frac{1}{2}}

\def\lim{\rightarrow}

\def\apr{a^{\prime}(u)}
\def\bpr{b^{\prime}(v)}

\def\bu{\bar{u}}

\def\bv{\bar{v}}

\def\cA{{\cal A}}
\def\cB{{\cal B}}
\def\cC{{\cal C}}
\def\cD{{\cal D}}
\def\cE{{\cal E}}
\def\cF{{\cal F}}
\def\cG{{\cal G}}

\def\hep{\hat{\epsilon}}

\def\ha{\hat{a}}

\def\hA{\hat{A}}
\def\hB{\hat{B}}
\def\hv{\hat{v}}
\def\hV{\hat{V}}
\def\hO{\hat{O}}

\def\tA{\tilde{A}}

\begin{document}
\raggedbottom

\title{{\bf Multi - exponential  models of \\ (1+1)-dimensional
          dilaton  gravity and \\ Toda - Liouville integrable models}}

\author{V. de Alfaro \thanks{vda@to.infn.it}\\{\small \it $^*$ Dip. Fisica
    Teorica, INFN, Accademia Scienze; v.Giuria 1, 10125 Torino IT}\\
A.T.~Filippov \thanks{Alexandre.Filippov@jinr.ru}~ \\
{\small \it {$^\dagger$
 Joint Institute for Nuclear Research, Dubna, Moscow
Region RU-141980} }}

\maketitle

\begin{abstract}

 The general properties  of a class of two-dimensional dilaton gravity (DG)
theories with multi-exponential potentials are studied and a subclass of
these theories, in which the equations of motion reduce to Toda and
Liouville equations, is treated in detail. A combination of parameters of
the equations should satisfy a certain constraint that is identified and
solved for the general multi-exponential model. From the constraint it
follows that in DG theories the integrable Toda equations, generally,
cannot appear without accompanying Liouville equations.

 The most difficult problem in the two-dimensional Toda - Liouville DG is
 to solve the energy and momentum constraints. We discuss this problem using
 the  simplest examples and identify the main obstacles to finding its
 analytic solution.
   Then we consider a subclass of integrable two-dimensional theories, in
   which scalar matter fields satisfy the Toda equations while  the
 two-dimensional metric is trivial; the simplest case is considered in some
 detail, and on this example we outline how the general solution can be
 obtained.

  We also show how the wave-like solutions of the general
  Toda - Liouville systems can be simply derived. In the dilaton gravity
  theory, these solutions describe nonlinear waves coupled to gravity as
  well as static states and cosmologies.
  For static states and cosmologies we propose and study a more general
  one-dimensional Toda - Liouville model typically emerging in
  one-dimensional reductions of higher-dimensional gravity and
  supergravity theories.
   A special attention is paid in this paper to
  making the analytic structure of the solutions of the Toda equations as
  simple and transparent as possible, with the aim to gain a better
  understanding of realistic theories reduced to dimensions 1+1
    and 1+0 or 0+1.

\end{abstract}

\section{Introduction}

The theories of $(1+1)-$dimensional dilaton gravity coupled to scalar
matter fields are known to be reliable models for some aspects of
higher-dimensional black holes, cosmological models and waves. The
connection between higher and lower dimensions was demonstrated in
different contexts of gravity and string theory and, in several cases, has
allowed finding the general solution or special classes of solutions in
high-dimensional theories \footnote{See, e.g.,  \cite{BZ}-\cite{Venezia}
for a more detailed discussion of this connection, references, and
solution of some integrable two-dimensional and one-dimensional models of
dilaton gravity. }. A generic example is the spherically symmetric gravity
coupled to Abelian gauge fields and scalar matter fields. It exactly
reduces to a (1+1)-dimensional dilaton gravity and can be explicitly
solved if the scalar fields are constants  independent of
  the coordinates\footnote{This is not possible for arbitrary dependence
  of the potentials on the scalar fields, as it
 will be clear in a moment.}.
  These solutions can describe interesting physical objects --
spherical static black holes and simplest cosmologies. However, when the
scalar matter fields, which presumably play a significant cosmological
role, are nontrivial, not many exact analytical solutions of
high-dimensional theories are known\footnote{See, e.g.,
         \cite{CGHS}, \cite{NKS}, \cite{ATF4}, \cite{VDA1}-\cite{ATF5};
         a review and further references can be found in \cite{Strobl},
  \cite{Kummer} and \cite{ATF5}.}.
  Correspondingly, the two-dimensional models of DG that
  nontrivially couple to scalar matter are usually not integrable.

  To construct integrable models of this sort one usually must apply
 serious approximations, in other words, deform the original
 two-dimensional model obtained by direct dimensional reductions of
 a realistic higher-dimensio\-nal theory.
 Nevertheless, the deformed models can qualitatively describe certain
 physically interesting solutions of higher-dimensional gravity or
 supergravity theories related to the low-energy limit of superstring
  theories.
 We note that several
 important four-dimensional space-times with symmetries defined by two
 commuting Killing vectors may also be described by two-dimensional
 models of dilaton gravity coupled to scalar matter. For example,
  cylindrical gravitational waves can be described by a $(1+1)-$dimen\-sio\-nal
  dilaton gravity   coupled to one scalar field
  \cite{Einstein}-\cite{Chandra1}, \cite{ATF3}.
 The stationary axially symmetric pure gravity (\cite{Ernst}, \cite{NKS})
  is equivalent to a $(0+2)-$dimensional dilaton gravity coupled to one scalar
field. Similar but more general dilaton gravity models were also obtained
in string theory. Some of them can be solved by using modern mathematical
 methods developed in
 soliton theory (see e.g. \cite{BZ},
\cite{Maison}, \cite{NKS}, \cite{Alekseev}).
 Note also that the theories in
 dimension 1+0 (cosmologies) and 0+1 (static states and, in particular,
 black holes) may be integrable in spite of the fact that their 1+1
 dimensional `parent' theory is not integrable without a deformation (see
 [23] and an example given in this paper).

  In our previous work (see, e.g., \cite{ATF2} - \cite{ATF5} and references
 therein) we constructed and studied some explicitly integrable models
 based on the Liouville equation.
 Recently, we attempted to find solutions of some realistic
     two-dimensional dilaton gravity models (derived from higher-dimensional
      gravity theories by dimensional reduction) using a generalized
 separation of variables introduced in \cite{W1}, \cite{ATF3}.
 These attempts showed that seemingly
natural ansatzes for the structure of the separation, which proved a
  success in previously studied integrable models, do not give
interesting enough solutions (`zero' approximation of a perturbation
theory) in realistic nonintegrable models. Thus an investigation of more
complex dilaton gravity models, which are based on the two dimensional
 Toda  chains, was initiated in \cite{ATF6}.

 At first sight
 it seems that it should be not difficult to find a potential in DG
 theory that will give integrable Toda equations of motion. However in
 reality it is not as simple as that, and the Toda theory may only
 emerge in company with a Liouville theory (this was mentioned in
 footnote in ref.~\cite{ATF6}). In fact, even the $N-$Liouville theory
 satisfies the same constraint.  It was known to the authors of \cite{ATF5}
  and \cite{ATF6} since long time
  but the meaning of this fact was not clearly understood.

In this paper we first introduce the general {\bf multi-exponential} DG
and present the equations of motion in a form that resembles the Toda
equations. In addition to the equations, in the DG theory one should
satisfy two extra equations which in General Relativity are called
 {\bf the energy and momentum constraints}.
 In the $N-$Liouville theory these
 constraints were explicitly solved but in the general case solving the
 constraints is a much more difficult problem which we discuss in Section~4.

Section~3 is devoted to the problem of reconstructing the dilaton gravity
from the `one-exponen\-tial' form of the equation of motion
 \be
 \label{001}
\d_u\d_v \,x_m\,=\,g_m\exp{\sum_n\,A_{mn}\,x_n}\,.
 \ee
This amounts to finding the matrix $\ha$ satisfying the matrix
equation\footnote{We call it the A-equation.} $\ha^T \hep\ha=\hA$ ($\hep$
is a diagonal matrix to be introduced later). Evidently, this equation may
have many solutions for a fixed matrix $\hA$ (e.g., if $\ha$ is a
solution, then $\hO\ha$, where $\hO^T\hep\hO=1$, is also a solution). The
 important fact is however that {\bf the solution is not possible for an
 arbitrary} symmetric matrix $\hA^T=\hA$. In Section~3 we establish the
class of `solvable' matrices $\hA$ (satisfying the A-condition) and
introduce a recursive procedure in order to find all possible solutions
for any matrix satisfying the $A-$condition.
   In Appendix~2 we give the general solution of the A-condition for
   the matrix $A_{mn}$ being the direct sum of a diagonal $L \times L$
   matrix and of an arbitrary symmetric matrix ($N$-Liouville plus
   multi-exponential model).

 The Cartan matrices for simple Lie groups do not satisfy the A-condition
 and thus {\bf the generic DG cannot be reduced to the
 Toda equations}.\footnote{
  Due to the A-equation, the $A_{mn}$ in Eq.(\ref{001}) must
  be symmetric. When the Cartan matrices are non-symmetric,
 Eq.(\ref{001}) depends on the symmetrized matrices
 (see Appendix~1).}
       However, adding at least one Liouville equation to the Toda system (Toda
- Liouville System, or TL) solves this constraint  and in Section~4 we
briefly introduce the simplest form of solution of TLS in the case of the
 $\cA_n$ Cartan matrices.
    In addition to these standard solutions, we construct the wave-like
    solutions similar to ones earlier derived in the $N$-Liouville model.
    For these solutions the energy-momentum constraints are easily
    satisfied. In Appendix~3 we show that the form of the constraints
    for the general solution
    in the simplest $\cA_1 \bigoplus \cA_2$ model is the same, but this
    does not help to solve them (this follows from the result of
    Appendix~2).

  In Section~5 we turn to a simpler class of Toda - based DG models that can
  be completely solved (the energy-momentum constraints included).
  If we suppose that the potential $V$ is
independent of the dilaton $\f$ (i.e., $V_{\f}=0$), then the metric is
flat (in the Weyl frame) and the constraints can be solved once we solve
the Toda equations, which in this case need not be accompanied by the
Liouville equations.  This model is, in fact, a far going generalization
of the well known CGHS model and can be solved directly and explicitly
(although the properties of the solution are much, much more complex than
their CGHS counterpart).

  Section~6 is devoted to the investigation of realistic one-dimensional
 TL models of cosmologies and static states (e.g. black holes) that can be
derived from higher dimensional gravity or supergravity. Models of this
sort have been known for more than a decade but it seems that the need of
the Toda - Liouville connection was not realized. We give here a complete
treatment of this connection by a proper generalization of the
A-condition.

 Finally, in Section~7 we summarize our results in a more dogmatic form,
 emphasizing unsolved problems and possible applications
 to black holes, cosmologies and waves.

\section{Multi - exponential model of (1+1)-dimensional dilaton gravity \\
  minimally coupled to scalar matter fields.}

The effective Lagrangian of the (1+1)-dimensional dilaton gravity coupled
to scalar fields $\psi_n$ obtainable by dimensional reductions of a
higher-dimensional spherically symmetric (super)gravity can usually be
(locally) transformed to the form:
\be
 {\cal L}^{(2)} = \sqrt{-g}\left[ \f R(g) + V(\f,\psi) +
 \sum_{m,n} Z_{mn}(\f,\psi)\, g^{ij} \, {\dif}_i \psi_m \, {\dif}_j \psi_n \right] \,
\label{7}
\ee
(see \cite{ATF2} - \cite{ATF5} for a detailed motivation and examples).
 In Eq.(\ref{7}), $g_{ij}(x^0,x^1)$ is the (1+1)-dimensional metric with signature
 (-1,1), $g \equiv {\rm det}(g_{ij})$, $R$ is the Ricci curvature of
 the two-dimensional space-time with the metric
\be
ds^2=g_{ij}\, dx^i \, dx^j \, , \,\,\,\,\,\, i,j = 0,1 \, .
\label{4}
\ee
The effective potentials $V$ and $Z_{mn}$ depend on the dilaton
 $\f(x^0,x^1)$ and on $N-2$ scalar fields $\psi_n(x^0,x^1)$
 (we note that the matrix $Z_{mn}$ should be negative definite
 to exclude the so called `phantom' fields).
 They may depend on other parameters characterizing the parent
higher-dimensional theory (e.g., on charges introduced in solving the
equations for the Abelian fields).
         Here we consider the `minimal'
 kinetic terms with  diagonal and constant $Z$-potentials,
 $Z_{mn}(\f, \psi) = \delta_{mn} Z_n$\footnote{
 In Section~6 we add to the Lagrangian (\ref{7}) with minimal $\psi$-coupling
 certain fields $\sigma$ whose $Z$-potentials depend on $\f$
      and $\psi$ but are independent of $\sigma$, see, e.g., \cite{ATF5}.}.
 This approximation
 excludes the important class of the sigma - model - like
 scalar matter discussed, e.g., in \cite{Venezia}; such models can be
 integrable if $V \equiv 0$ and the
 $Z_{mn}(\f,\psi)$ satisfy certain rather
 stringent conditions.
 In (\ref{7}) we also used the Weyl transformation to eliminate
 the gradient term for the dilaton.
 To simplify derivations, we write the equations of motion in the
 light-cone metric,   $$ds^2 = -4f(u, v) \, du \, dv \, .$$
 By first varying the Lagrangian in generic coordinates and then passing
 to the light-cone coordinates we obtain the equations of motion
  ($Z_n$ are constants!)
 \be
 \p_u \p_v \f+f\, V(\f,\psi) = 0,
 \label{F.15}
 \ee
  \be
  f \p_i ({{\p_i \f} / f }) \, = \sum Z_n \,(\p_i \psi_n)^2\, ,
\,\,\,\,\,\,\,\,\, i=u,v \, .
 \label{F.17}
 \ee
 \be
 2Z_n \,\p_u \p_v\, \psi_n + f\, V_{\psi_n}(\f,\psi)= 0 \, ,
 \label{F.16}
 \ee
 \be
 \p_u\p_v\ln |f| + f V_{\f}(\f,\psi) = 0 \, ,
 \label{F.18}
 \ee
 where $V_{\f} \equiv \p_{\f} V$, $V_{\psi_n} \equiv \p_{\psi_n} V$.
 These equations  are not independent. Actually, (\ref{F.18}) follows
from (\ref{F.15}) $-$  (\ref{F.16}). Alternatively, if  (\ref{F.15}),
(\ref{F.17}), and (\ref{F.18}) are satisfied, one of the equations
(\ref{F.16}) is also satisfied. Note that the equations may have the
solution with $\psi_n = \psi_n^{(0)} = \textrm{const}$ only if
 $V_{\psi_n}(\f , \psi_n^{(0)}) \equiv 0$.

The higher-dimensional origin of the Lagrangian (\ref{7}) suggests that
the potential is the sum of exponentials of linear combinations of
 the scalar fields
and of the dilaton $\f$. \footnote{Actually, the potential $V$ usually
contains terms
       non exponentially depending on $\f$ (e.g., linear in $\f$),
       and then the exponentiation of $\f$ is only an approximation,
       see the discussion in \cite{ATF5}.}
     In our previous work \cite{ATF5}
 we studied the constrained Liouville model, in which
the system of equations of motion (\ref{F.15}), (\ref{F.16}) and
(\ref{F.18}) is equivalent to the system of independent Liouville
equations for the
 linear combinations of  fields $q_n \equiv F + q_n^{(0)}$, where
 $F \equiv \ln|f|$. The easily derived solutions of these equations should
 satisfy  the constraints (\ref{F.17}), which was the most difficult part of
 the problem. The solution of the whole problem revealed an interesting
 structure of the moduli space of the solutions that allowed us to easily
 identify static, cosmological and wave-like solutions and effectively
 embed these essentially one-dimensional (in some broad sense) solutions
 into the set of all two-dimensional solutions and study their analytic
 and asymptotic properties.

 Here we propose a natural generalization of the Liouville model to the
 model in which the fields are described by the Toda equations (or by
 nonintegrable deformations of them). To demonstrate that the model shares
 many properties with the Liouville one and to simplify a transition from the
          integrable models to nonintegrable theories we suggest a different
 representation of the Toda solutions which is not directly related to
 their group - theoretical background (Section~4).

  Consider the theory defined by the Lagrangian (\ref{7})
 with the potential
 \be
 V = \sum_{n=1}^N 2g_n \exp{q_n^{(0)}} \, , \qquad
 q_n^{(0)} \equiv a_n \f + \sum_{m=3}^{N} \psi_m a_{mn} \, ,
 \label{8}
 \ee
 and with $Z_{mn}(\f, \psi) = -\delta_{mn}$.
  In what follows we also use the fields
 \be
 q_n \equiv  F + q_n^{(0)} \equiv \sum_{m=1}^{N} \psi_m a_{mn} \, ,
 \label{9aa}
 \ee
where $\psi_1 + \psi_2 \equiv \ln{|f|} \equiv F$,
$\psi_1 - \psi_2 \equiv \f$ and hence
$a_{1n} = 1 + a_n$, $a_{2n} = 1 - a_n$.

 Rewriting the equations of motion in terms of $\psi_n$
 we find that  Eqs.~(\ref{F.15}) - (\ref{F.18}) are equivalent to
 $N$ equations of motion for $N$ functions $\psi_n$
 ($\varepsilon$ is the sign of the metric $f$),
\be
 \p_u \p_v \psi_n  =
  \varepsilon  \sum_{m=1}^{N} \epsilon_n a_{nm} g_m \exp{(q_m)} \quad
 ( \epsilon_1 = -1, \,\,\, \epsilon_n = +1 \,\, {\rm if} \,\, n \geq 2 \,)\, ,
\label{10}
\ee
and two constraints,
\be
 C_i \equiv \p_i^2 \f + \sum_{n=1}^N  \epsilon_n  (\p_i \psi_n)^2 =  0,
 \,\,\,\,\,\,\, i=u,v \, .
 \label{11}
\ee
 With arbitrary parameters $a_{nm}$, these equations of motion are not
 integrable. However,
 as proposed in \cite{A2} - \cite{ATF1}, \cite{ATF2} \cite{ATF5},
 Eqs.(\ref{10}) are integrable and the constraints (\ref{11}) can be solved
 if the $N$-component  vectors $v_n \equiv (a_{mn})$ are pseudo-orthogonal.

 Now, consider more general nondegenerate matrices $a_{mn}$ and
       define the new scalar fields $x_n$:
\be
       x_n  \equiv  \sum_{m=1}^{N} a_{nm}^{-1} \epsilon_m \psi_m  \, ,
   \,\,\,\,\,\,\,\,\,\,\,\,\,\,\,
       \psi_n  \equiv  \sum_{m=1}^{N} \epsilon_n a_{nm} x_m  \, .
 \label{12}
\ee
 In terms of these fields, Eqs.(\ref{10})  read as
\be
  \p_u \p_v x_m  \equiv  \varepsilon g_m
     \exp ({\sum_{k,n=1}^{N}  \epsilon_n a_{nm} a_{nk} x_k}\,)
     \equiv  \varepsilon g_m \exp ({\sum_{k=1}^{N} A_{mk} x_k}\,)  \, ,
 \label{13}
 \ee
 and we see that the symmetric matrix
   \be
  \hA \,\equiv \,\ha^T \,\hep \,\ha \, , \qquad \ep_{mn} \equiv
  \ep_m \ \delta_{mn} \,,
\label{14}
    \ee
     defines the main properties of the model.

 If $\hA$ is a diagonal matrix  we return to the $N$-Liouville model.
 If $\hA$ were the Cartan matrix of
 a simple Lie algebra, the system (\ref{13})
 would coincide with the corresponding
 Toda system, which is integrable and can be more or less explicitly
         solved (see, e.g., \cite{Leznov}, \cite{Saveliev} ).
  However, in Section~3 we show that the Cartan matrices
         of the simple Lie algebras (symmetrized when necessary)
          cannot be represented in the form (\ref{14}).
  Nevertheless, a very simple extension of the Toda equations obtained by adding one or
  more Liouville equations can resolve this problem. In fact, a symmetric
  matrix $A_{mn}$ that is the direct sum of
  a diagonal $L\times L$-matrix $\gamma_n^{-1} \delta_{mn}$
  and of an arbitrary symmetric matrix $\bar{A}_{mn}$,
  can be represented in form (\ref{14}) if the sum of $\gamma_n^{-1}$
  is a certain function of the matrix elements $\bar{A}_{mn}$ .
  If $\bar{A}_{mn}$ is a Cartan matrix, the system (\ref{13})
  thus reduces to $L$ independent Liouville (Toda $\cA_1$) equations and the
  higher-rank Toda system (Toda - Liouville system, or, TLS).

 The solution of TLS can be derived in several ways. The most general one
 is provided by the group-theoretical construction described in
 \cite{Leznov}, \cite{Saveliev}. Here, in Section~4 we outline an
 analytical method directly applicable to solving $\cA_N$ TLS proposed in
 \cite{ATF6}. However, solving the equations of motion is not the whole
 story.
  Once the equations are solved, their solutions must be
 constrained to satisfy the zero energy-momentum conditions (\ref{11})
 that in terms of $x_n$ are:
 \be
 -C_i \, \equiv \, 2\sum_{n=1}^N \ \p_i^2 x_n \, - \,\sum_{n,m =1}^N \ \p_i x_m \
 A_{mn} \, \p_i x_n \, = \,0 \,, \quad i=u,v \, .
 \label{15}
 \ee
 In the $N$-Liouville model the most difficult
 problem was to satisfy the constraints (\ref{15}) but this problem was
 eventually solved. In the general nonintegrable case
            of an arbitrary matrix
 $\hA$, we do not know even how to approach this problem.
 The Toda - Liouville case is discussed below\footnote{
 In Section~5 we introduce a simplified DG Toda
 model that can be completely solved, including the constraints.}.

 To study the general properties of the solutions
 of equations (\ref{13}) and of the constraints (\ref{15})
 we first rewrite the general equations in a form that is particularly
 useful for the Toda - Liouville systems. Introducing the notation
  \be
 X_n \equiv \exp (-\half A_{nn} x_n) \ , \,\,\,\,\,\,
  \Delta_2 (X) \equiv X\ \p_u \p_v X - \p_u X \ \p_v X , \,\,\,\,\,\,
  \alpha_{mn} \equiv -2 A_{mn} / A_{nn} \,,
 \label{16}
 \ee
 it is easy to rewrite Eqs.(\ref{13}) in the form:
 \be
 \Delta_2 (X_n) =
  -\half \varepsilon \ g_n A_{nn} \prod_{m \neq n} X_m^{\alpha_{nm}} \, .
 \label{17}
 \ee
 The multiplier $|-\half \varepsilon \ g_n A_{nn}|$ can be removed by using
 the transformation $x_n \mapsto x_n + \delta_n$ and the final (standard)
 form of the equations of motion is
 \be
 \Delta_2 (X_n) =
  \varepsilon_n  \prod_{m \neq n} X_m^{\alpha_{nm}} \, , \qquad
  \varepsilon_n \equiv \pm 1 .
 \label{18}
 \ee

 These equations are in general not integrable. However, when $A_{mn}$
     are Toda plus Liouville matrices, they simplify to integrable equations
 (see \cite{Leznov}). The Liouville part is diagonal while the Toda part
 is non-diagonal.
   For example, for the Cartan matrix of $\cA_N$, only the
   near-diagonal elements of the matrix $\alpha_{mn}$
   are nonvanishing, $\alpha_{n+1,n-1} = \alpha_{n-1,n+1} = 1$.
 This allows one to solve Eq.(\ref{18}) for any $N$.
 The parameters $\alpha_{mn}$ are invariant w.r.t. the
 transformations $x_n \mapsto \lambda_n x_n + \delta_n$.
   This means that the non-symmetric Cartan matrices of
 $\cB_N$, $\cC_N$, $\cG_2$, and $\cF_4$ can be symmetrized while not
 changing the equations. In this sense, the
 $\alpha_{mn}$ are the fundamental parameters of the equations of
 motion. From this point of view, the characteristic property of the
 Cartan matrices is the simplicity of Eqs.(\ref{18}) which allow one
 to solve them by a generalization of separation of variables.
 As is well known, when $A_{mn}$ is the Cartan matrix of any simple
  algebra, this procedure gives
 the exact general solution (see \cite{Leznov}). In  Section~4 we
 show how to construct the exact general solution for the $\cA_N$ Toda system
 and write a convenient representation for the general solution that
 differs from the standard one given in \cite{Leznov}.

 Unfortunately, as we emphasized above, solving equations (\ref{18})
 is not sufficient for finding the solution of the whole problem.
 We also must solve the constraints (\ref{15}), and this is a much more
 difficult task. In our previous papers we succeeded in solving the
 constraints of the $N$-Liouville theory. So, let us try to formulate the
 problem of the constraints in the Toda - Liouville case as close as possible
 to the $N$-Liouville case.
    First, it is not difficult to show that
    $\d_v\,C_u\,=\,\d_u\,C_v\,=\,0$ and thus
    $C_u = C_u (u)$, $C_v = C_v (v)$ as in the Liouville case.
    To prove this one should differentiate (\ref{15}) and use (\ref{13})
    to get rid of $\p_u \p_v x_m$ and $\p_u \p_v x_n$.

 Up to now we considered an arbitrary symmetric matrix $\hA$. At this point
 we should use a more detailed information about $A_{mn}$ and about the
 structure of the solution.
  To see whether the constraints can be solved we first rewrite them
  in terms of $X_n$ and then consider
  the Toda - Liouville matrices and
  the explicit solutions of the equations.
 It is not difficult to see that the constraints (\ref{15}) can be written
 in the form ($i=u$ or $i=v$ and the prime denotes $\partial_i$):
  \be
  \label{f30}
  {1\over 4} C_i =
 \sum_{n=1}^N \gamma_n{X_n^{''}\over X_n} \,+\, \sum_{m<n}^N
 {2A_{mn}\over A_mA_n}\, {X_m^{'}\over X_m}\,
 {X_n^{'}\over X_n}\,, \quad   \gamma_n\equiv A_n^{-1}\,.
\ee
 The first term looks exactly as in the case of the $N-$Liouville model.
 However, in the Liouville case we also knew that
 \be
 \label{f35}
 \d_u\biggl(X_n^{-1} \,\d_v^2 X_n\biggr) = \,0,
 \quad \d_v\biggl(X_n^{-1} \,\d_u^2 X_n\biggr)=\,0,
 \ee
which is not true in the general case. Moreover, the first and the second
terms in r.h.s. of Eq.(\ref{f30}) are in general not
 functions of a single variable (above we have only proved
  that in general $C_u=C_u(u)$ and $C_v=C_v(v)$).

Nevertheless, let us try to push the analogy with the Liouville case as
far as possible, at least in the integrable Toda - Liouville case. Thus,
suppose that the first $N_1$ equations are the Toda ones and the remaining
$N_2=N-N_1$ equations are the Liouville ones. This means that
$A_{mn}=\tA_{mn} $ ($1\leq m,n\leq N_1$), where $\tA_{mn}$ is a Cartan
matrix while for $N_1+1\leq m,n\leq N$ we have $A_{mn}=\delta_{mn} A_n$.
Then the constraints split into the Toda and the Liouville parts
  ($X^{'}\equiv \partial_i X$):
 \be
 \label{f40}
 {1\over 4} C_i = \sum_{n=1}^{N_1}\,{1\over A_n}\,{X_n^{''} \over X_n} \,+\,
  \sum_{m<n}^{N_1}\, {2A_{mn}\over A_mA_n}\, {X_m^{'}\over X_m}\,
  {X_n^{'}\over X_n}\, + \,\sum_{n=N_1+1}^N\, \gamma_n {Y_n^{''} \over Y_n}\,.
 \ee
  They are significantly different: first, because
  the Liouville solutions $Y_n$ for $n\geq N_1+1$
 satisfy the second order differential equation while the Toda solutions
 $X_n$ satisfy higher order ones (see Section 4).
  In the general $\cA_N$ Toda case $X_1$ can be written as
 \be
 \label{f45}
  X_1=\sum_{i=1}^{N+1}\,a_i(u) \,b_i(v) \,,
  \quad X_2 = \varepsilon_1 \De_2(X_1),
 \ee
     while in the Liouville case the solution is simply the sum
     of two terms (see Section 4).
     Moreover, for the Liouville solution $Y(u,v)$ we have
 \be
 \label{f55}
 Y^{-1} \d_u^2 Y = {a_1^{''}(u)\over a_1(u)}
 ={a_2^{''}(u)\over a_(u)} \,,
 \qquad
 Y^{-1} \d_v^2 Y = {b_1^{''}(v) \over b_1(v)}
 = {b_2^{''}(v) \over b_2(v)} \,,
  \ee
while in the Toda case everything is much more complex.

To understand better this fact we consider the case $N_1=2$, $N=3$ with
$A_{mn} (1\leq m,n\leq 2)$ being the $\cA_2-$ Cartan matrix and
$A_{3n}=\delta_{3n} A_3$.
 Using $A_1=A_2=2$, $A_{12}=A_{21}=-1$, we find
 \be
 \label{f60}
 {1\over 2} C_i\,=\, \biggl({X_1^{''} \over X_1} +{X_2^{''}\over X_2} -
 {X_1^{'}\over X_1}\cdot {X_2^{'}\over X_2}\biggr) \,-\,
  4{Y_3^{''}\over Y_3} \, = 0
 \ee
 where $X_2 = \varepsilon_1 \De_2(X_1)$, $\varepsilon_1 = \pm 1$,
 and $Y_3$ is the Liouville solution
 (note that according to the constraint on
 $A_{ij}$, considered in next Section,
 we have in this case $\gamma_3=A_3^{-1}=-2$,
 as can be seen from Eq.(\ref{b10}) below).
 Although we know
that $Y_3^{''}/Y_3$ and $C_i$ are functions of one variable, we do not
have at the moment simple and explicit expressions for $C_i$.
 Indeed, using (\ref{f45}) it is not difficult to find that
 \be
 \label{f65}
  \d_v (X_1^{-1} \d_u^2 X_1) = \,
   = \,{1\over 2} \biggl(\sum_{j=1}^3 a_j\,b_j \biggr)^{-2}\,\sum_{i,j}
  W^{'}[a_i,a_j]\, \,W[b_i,b_j] \,\not=\,0\,.
  \ee
 So, we should first write the explicit expression for $X_2(u,v)$ in
 terms of $a,\,b,$ and then derive the complete first term in $C_i$.
 We construct the
 solutions of the $\cA_2 \bigoplus \cA_1$ constraints in Section 4.

\section{Solving $\ha^T\,\hep\,\ha\,=\,\hA$}
 In this section we show how to solve Eq.(\ref{14}) for the matrix $\ha$ in
 the standard DG. This is possible if and only if $\hA$ satisfies certain
 conditions, which we explicitly derive.
  First, $\det \hat{A} = -\det \hat{a}^2<0$.
   This restricts the matrices $\hA$ of even order but is not so severe a
 restriction for the odd order matrices. In fact, we can then change sign
 of $\hA$ and of all the variables $x_n$ and the only effect will be
 that all $\varepsilon_n$ in Eq.(\ref{18}) change sign. If these signs are
 unimportant and the two systems of equations may be considered as
 equivalent, the restriction does not work. As the
 determinants of all (symmetrized) Cartan matrices for simple groups are
 positive (and their eigenvalues are positive), it follows
 that the even-order Cartan matrices do not satisfy this restriction.
 A more severe restriction is related to the special structure of
 the matrices $a_{mn}$ in (\ref{9aa}). In consequence, the matrix $\hA$
 must satisfy one equation that we derive and explicitly solve below.
    In this section we consider the standard DG and in Section 6 we analyze
  in the same approach a somewhat different one-dimensional dilaton gravity
  which can be met in cosmological models.

 Let us now take the general $N \times N$ matrix $\ha$ of DG, with the only
 restriction: $a_{1n} = 1+a_n$ and $a_{2n} = 1-a_n \,$.
      The equations defining $a_{mn}$ in terms of $A_{mn}$ are
\be
\label{1}
-2(a_m+a_n) \,+V_m \cdot V_n \,=\, A_{mn} \,,
 \qquad    -4a_n\,=\,A_n-V_n^2\,,
 \qquad  m,n=1,...,N
\ee
 where we introduced the
 notation  $V_n\,\equiv\,(\,a_{3n},...,a_{Nn})$.
 As it follows from (\ref{1}),
 our $N$ vectors $V_i$ in the $(N-2)-$dimensional space have $N(N-2)$
 components and satisfy  $N(N-1)/2$ equations:
 \be
 \label{1a}
 (V_m-V_n)^2 \,=\,A_m+A_n-2A_{mn}\,, \qquad m>n,\,\,\,m,n=1,...,N.
 \ee
 These equations are invariant under $(N-2)\,(N-3)/2$ rotations of the
 $(N-2)-$ dimensional space and under ${N-2}$ translations.
  It follows that the vectors $V_m$ in fact depend on $(N-2)\,(N+1)/2$
  invariant parameters and the number of
   equations minus the number of parameters is equal to one.
   Therefore, one of the equations should give a relation between the parameters.

 It is possible to give a more constructive approach directly
  utilizing  the invariant equations that
  follow from the equations $(\ref{1a})$.
  Define $v_k\equiv V_k-V_1$
 ($k=2, ...,N$); then, from $(\ref{1a})$ we have:
\bdm
  v_k^2\,\equiv\,(V_k-V_1)^2 \,=\,A_1+A_k-2A_{1k} \,\equiv \,
  \tilde{A}_{1k}\, ,
\edm
  \bdm
   (v_k-v_l)^2 \,\equiv \,
   \tilde{A}_{1k} \,+\, \tilde{A}_{1l} \,-\, 2v_k \cdot v_l \, ,
   \qquad k>l; \,\,\,\, k,l=2,...,N\, .
  \edm
  Thus the general invariant equations for $v_k$ can be written:
 \be
 v_k \cdot v_l\,=\, A_1-A_{1k}-A_{1l}+A_{kl} \,, \qquad k \geq l \,.
 \label{2a}
 \ee
  As these equations are valid also for $l=k$ we have $N(N-1)/2$
  equations for the same number of the invariant parameters
  $v_k \cdot v_l$.
  But, of course, there is one relation between these parameters because
  there exist a linear relation between $N-1$ vectors $v_k$ in the $(N-2)-$
  dimensional space. For example, $v_N^2$ can be expressed in terms of the
  remaining parameters $v_2^2, ..., v_{N-1}^2$ and $v_k \cdot v_l$, $k>l$
  (their number is $(N-2)(N+1)/2$, as above).
  As the equations for $v_k$ express $v_k \cdot v_l$ in terms of the matrix
  elements $A_{kl}$, we thus can derive
  the necessary relation between $A_{kl}$ (e.g., an expression
  of $A_1\equiv A_{11}$ in terms of the remaining matrix elements).

   Using the vectors $v_k$ we can give an explicit construction of the
 solutions and derive the constraint on the matrix elements $A_{mn}$.
 The construction of the solution of the equations for
 $a_{mn}$ can be given as follows. It is not difficult to understand
 that we only need to find the unit vectors,
 \be
 \label{ex1}
 \hv_k \equiv {v_k / |v_k|}\,=\,v_k\,\tilde{A}_{1k}^{-1/2} \,,
 \ee
 in any fixed coordinate system in the $(N-2)-$ dimensional space. Then we
 can reconstruct the general solution by applying to $\hv_k$ rotations
 and translations (i.e. choosing arbitrary $a_{n1}$, $n=3,...,N$).
 Let us introduce the temporary notation
\be
 c_{kl}\equiv \cos\theta_{kl} \equiv
   \hv_k \cdot \hv_l \,=\,
 (A_1-A_{1k}-A_{1l}+A_{kl}) \, (\tilde{A}_{1k} \, \tilde{A}_{1l})^{-1/2}.
\ee
As  $v_k=(a_{3k}-a_{31}, ..., a_{Nk}-a_{n1}\,)$, we denote
$\alpha_{nk}\equiv (a_{nk}-a_{n1})/|v_k|$ and thus $\hv_k=(\alpha_{3k},
..., \alpha_{Nk})$. Choosing the coordinate system in which
 $\hv_2=(1,0,..0)$ we see that
 $\alpha_{3k}=c_{k2}\equiv \cos\theta_{2k}$ and $\hv_3$
 can be chosen with two nonvanishing components,
\be
 \hv_3 =(c_{23}, s_{23},0,...,0)\,,
\ee
 where $s_{23} \equiv \sin\theta_{23}$ and in general
 $s_{kl}=\sin \theta_{kl}$.
  The further invariant parameters $\alpha_{nk}$ can be derived recursively.
  The vectors $\hv_k,...,\hv_N$ for $k\geq 4$ are constructed as follows.
  We take
  $\alpha_{3k}=c_{2k}$, $\alpha_{nk}=0$ if $k\leq N-2$ and $n\geq k+2$.
   Then
\be
\hv_k=(c_{2k},\alpha_{4k}, \alpha_{5k}, ...,\alpha_{(k+1)k}, 0, 0...)
\ee
 and the parameters $\alpha_{nk}$ can be
 recursively derived from the relations ($k\geq 4$)
\be
\sum_{n=4}^{l+1}\,\alpha_{nk}\alpha_{nl}=c_{kl}-c_{k2}c_{l2}\,;
 \quad k>l , \qquad \qquad
\sum_{n=4}^{k+1} \,\alpha_{nk}^2  \,=\, s_{k2}^2 ,\quad k\leq N-1\,.
\ee
 The normalization condition for
$\hv_{N}$ (not included in the above equations),
\be
\label{ex2}
\sum_{n=4}^N\, \alpha_{nN}^2 \,=\, s_{N2}^2 \,,
\ee
 then gives  a relation
 between the $c_{kl}$'s  (and thus between the $A_{ij}$'s).

 Using this solution we can find the expression for $A_1\equiv A_{11}$ in
 terms of $A_{kl}$. However, this derivation is rather
 awkward. It can be somewhat simplified if we consider simpler matrices
 $A_{kl}$ for which $A_{1k}=A_{k1}=0$, $k \neq 1$. Then one can find that
 the equation for $A_1$ is linear and
 thus has a  unique solution.
 Nevertheless it is not a good idea to derive the
 constraint on $A_{kl}$ in this rather indirect way. The linearity of the
 constraint in $A_1$ suggests that there exists a simple and general
 formula directly expressing $A_1$ in terms of the other elements $A_{kl}$.

 The simplest way to find $A_1$ in terms of the other $A_{ij}$ is the
 following: one of the vectors $v_2,\,v_3,\, ...,\,v_N$ must be given by a
 linear combination of $N-2$ other vectors. Suppose that
 \be
 \label{c20}
v_2=\sum_{p=3}^N \, v_p\,z_p\,.
 \ee
 Then we can find $z_p$ in terms of
 $A_{mn}$ by solving the equations
 \be
 \label{4a}
 v_p \cdot v_2\,=\,\sum_{q=3}^N\, (v_p \cdot v_q)\,z_q\,,\quad p=3,...,N\,.
 \ee
 The solution is given by $z_p=D_p/D$, where $D$ is the determinant
 of the matrix $(v_p \cdot v_q)$, and the $D_p$
 are the determinants of the same matrix but with the $p-$th column replaced
 by $(v_p \cdot v_2)$.

  Now it is clear that the expression of $v_2^2$ in terms of the solution
  of (\ref{4a}),
  \be
 \label{5a}
 v_2^2\,=\,\sum_{q=3}^N\, (v_2\cdot v_q) \, \,z_q\,
 =\,\sum_q\,(v_2 \cdot v_q) \,\cdot\,{D_q/ D}, \,
  \ee
 gives us the desired constraint on $A_{mn}$.
 Using (\ref{2a}) we rewrite it in the form
\be
\label{b10}
 (A_1+A_2-2A_{12})\,D\,=\,\sum_{p=3}^N\,
 (A_1+A_{p2}-A_{12}-A_{1p}\,)\, D_p\, ,
\ee
 where the determinants $D$ and $D_p$ should be expressed in terms of
  $A_{mn}$.  They evidently depend on $A_1$ linearly and thus
 Eq.(\ref{b10}) is at most quadratic in $A_1$.
 In fact, it is just linear. To prove this it is sufficient to show that
 \be
 \label{6a}
{dD \over dA_1}\,=\,\sum_{p=3}^N\, {dD_p \over dA_1}\,.
 \ee

 To simplify the proof we introduce the following temporal
 notation\footnote{
 Note that here $k,l = 2,3,...,N$ and $p,q,r = 3,...,N$.}
 \be
 \label{7a}
 D \equiv D(A_1) \equiv [C_3, C_4,...,C_N],\,\,\,\,\,\,\,\,\,\,\,
 D_q \equiv D_q(A_1) \equiv [C_3,...,C_{q-1},C_2, C_{q-1},...,C_N]\,,
 \ee
 where $C_k$ is the $k$-th column of the matrix
 $(v_p \cdot v_k)$, in particular, $C_2 \equiv (v_p \cdot v_2)$.
 To present differentiations in $A_1$ we additionally define the
 column $C_1$ all  elements of which
  are equal to one. In this notation
 we have (taking into account the simple dependence of $v_k \cdot v_l$
 on $A_1$, see (\ref{2a})):
 \be
 \label{8a}
 D^{\prime}(A_1) = \sum_{q=3}^N
 [C_3,...,C_{q-1},C_1, C_{q-1},...,C_N],\,\,\,,\,\,\,
 \ee
 \be
 \label{9a}
 D_q^{\prime}(A_1) = [C_3,...,C_{q-1},C_1, C_{q-1},...,C_N] \,+
 \, \sum_{r\neq q}^N
 [C_3,...,C_{r-1},C_1,C_{r-1},...,C_{q-1},C_2,C_{q-1} C_N] \,.
 \ee
Introducing the obvious notation, $D_{rq}(A_1)$, for the last
  determinants, we have
 \be
 \label{10a}
 D_q^{\prime}(A_1) = [C_3,...,C_{q-1},C_1, C_{q-1},...,C_N] \,+
 \, \sum_{r\neq q}^N D_{rq}(A_1) \,,
 \ee
 and thus (\ref{6a}) now has the form
 \be
 \label{11a}
 \sum_{q=3}^N D_q^{\prime}(A_1) = D^{\prime}(A_1) \,+\,
 \sum_{q=3}^N \sum_{r\neq q}^N D_{rq}(A_1) \,.
 \ee
 But the determinant $D_{rq}$ can be obtained from $D_{qr}$
 by an odd number of transpositions of the columns $C_1$, $C_2$
 and thus $D_{qr} = -D_{rq}$,
 which completes the proof.

 Now we can explicitly solve the constraint Eq.(\ref{b10}). Using
 the obvious relations
 \bdm
 D = D(0) \,+\, A_1 D^{\prime}(0) ,\,\,\,\,\,\,\,\,
 D_q = D_q(0) \,+\, A_1 D_q^{\prime}(0)
 \edm
 and the given above expressions for the determinants in terms
 of $A_{mn}$ one can write the general expression for $A_1$ in terms
 of the other matrix elements. We leave this as a simple exercise to
 the interested reader. Note only that the important case $A_{1n}=0$
 is somewhat simpler because then $(v_k \cdot v_l) - A_1 = A_{kl}$
 and thus $D(0) = \det(A_{pq})$, etc.
                              In Appendix we derive a
 beautiful solution of Eq.(\ref{b10}) for
 the models with  symmetric
 $N \times N$ matrices $A_{mn}$ having the form
 \bdm
 A_{mn} = \delta_{mn} A_n \,, \quad  1 \leq m,n \leq L  \, ;
 \qquad  A_{1n} = 0 \,, \quad  N \geq 2 \,.
 \edm

\section{Solution of the $\cA_N$ Toda system}
 The equations (\ref{18}) for the
 $\cA_N$-theory are extremely simple,
 \be
 \Delta_2 (X_n) =
  \varepsilon_n X_{n-1} X_{n+1} \, , \,\,\,\,\,\,\,\,
  X_0 \mapsto 1 \ , \,\,\,\, X_{N+1} \mapsto 1 , \,\,\,\,
 n = 1,...,N ,
 \label{19}
 \ee
 where $\varepsilon_n^2 =1$. As is well known, their solution can be
 reduced to solving just one higher-order equation for $X_1$
 by using the relation (see \cite{Leznov}):
 \be
 \Delta_2 (\Delta_n (X)) =
 \Delta_{n-1}(X) \ \Delta_{n+1}(X) \, , \,\,\,\,\,\,\,\,
 \Delta_1(X) \equiv X , \,\,\,\,\,\,  n \geq 2 \, ,
 \label{20}
 \ee
     where $\Delta_n (X)$ are determinants of the $n\times n$
     matrices $X_{km} \equiv \p_u^{\,k} \p_v^{\,m} X$
     ($1 \leq k,m \leq n$).
 Indeed, using Eqs.(\ref{19}), (\ref{20}) one can prove that for $n \geq 2$
 \be
  X_n = \Delta_n (X_1) \prod_{k=1}^{[n/2]} \varepsilon_{n+1-2k} \, ,
 \label{20a}
 \ee
 where the square brackets denote the integer part of $n/2$.
 Thus the condition $X_{N+1} = 1$ gives the equation for $X_1$,
 \be
 \Delta_{N+1} (X_1) = \prod_{k=1}^{[(N+1)/2]} \varepsilon_{N+2-2k} \,
  \equiv  \tilde{\varepsilon}_{N+1} \,=\, \pm 1 \, .
 \label{21}
 \ee
 This equation looks horrible but it is known to be exactly soluble
 by a special separation of variables, Eq.(\ref{f45}).
 We present its solution in
 a form that is equivalent to the standard one \cite{Leznov} but is
 more compact and more suitable for constructing effectively
 one-dimensional solutions, generalizing those studied in \cite{ATF5}.

     Let us start with the Liouville ($\cA_1$ Toda) equation
     $\Delta_2(X) = \tilde{\varepsilon}_2 \equiv \varepsilon_1$
 (see \cite{DPP}, \cite{Gervais}, \cite{Leznov}, \cite{ATF5}).
    Calculating the derivatives of $\Delta_2(X)$ in the variables $u$ and $v$
it is not difficult to prove Eqs.(\ref{f35}).
  It follows that there exist some `potentials' ${\cal{U}}(u)$,
${\cal{V}}(v)$ such that
\be
\label{23}
\p_u^2 X \,-\,{\cal U}(u) \, X\,=\,0 \, , \qquad \p_v^2 X \,-\,{\cal V}(v)
\, X\,=\,0 \,,
\ee
 and thus $X$ can be written in the `separated' form
 given in (\ref{f45}) with $N=1$
 where $a_i(u)$, $b_i(v)$ ($i = 1,2$) are linearly
 independent solutions of the equations (Eq.(\ref{f55})
 follows from this):
\be
\label{25}
a''_i(u) \, - \,{\cal U}(u) \, a_i(u) \, = \,0,
\qquad
b''_i(v)  \, - \, {\cal V}(v) \, b_i(v) \, = \, 0 \, .
\ee
 For $i=1$ these equations define the potentials for any choice of
  $a_1$, $b_1$, while
 $a_2$, $b_2$ then can be derived from the Wronskian first-order equations
\be
\label{26}
 W[a_1(u), a_2(u)] = w_a  \ , \,\,\,\,\,\,\,\,
 W[b_1(v), b_2(v)] = w_b \ , \,\,\,\,\,\,\,\,
     w_a \cdot w_b = \varepsilon_1 \ .
\ee
 We have repeated this well known derivation at some length because it is
 applicable to the $\cA_N$ Toda equation (\ref{21}).
 By similar derivations it can be shown that
 $X_1$ satisfies the equations
 \be
 \label{27}
 \p_u^{N+1} X + \sum_{n=0}^{N-1} {\cal U}_n(u) \ \p_u^n X = 0 \, ,
 \qquad
 \p_v^{N+1} X + \sum_{n=0}^{N-1} {\cal V}_n(v) \ \p_v^n X = 0 \, .
 \ee
 Thus the solution of (\ref{21}) can be written in the
 same `separated' form (\ref{f45}), where now
 $a_i(u)$, $b_i(v)$ ($i = 1,...,N+1$) satisfy the ordinary linear differential
 equations corresponding to (\ref{27}),
 with the constant Wronskians normalized by the conditions
 (one can choose any other normalization in
 which the product of the two Wronskians is the same):
 \be
 \label{28}
 W[a_1(u),..., a_{N+1}(u)] = w_a \ , \qquad
 W[b_1(v),..., b_{N+1}(v) ] = w_b \ , \qquad
 w_a \cdot w_b  = \tilde{\varepsilon}_{N+1} \ .
 \ee
 The potentials ${\cal U}_n(u)$ ${\cal V}_n(v)$
 can easily be expressed in terms of
 the arbitrary functions $a_i(u)$ and $b_i(v)$, $i=1,...,N$.
 To find the expressions one should differentiate the determinants
 (\ref{28}) to obtain the homogeneous differential equations for
 $a_{N+1}(u)$, $b_{N+1}(v)$. For example, for $N=2$:
 \be
 \label{28a}
 {\cal U}_1(u) = -(a_1 a'''_2 - a'''_1 a_2)/ W[a_1,a_2] , \qquad
 {\cal U}_0(u) = (a'_1 a'''_2 - a'''_1 a'_2)/ W[a_1,a_2] \ .
 \ee

 As an exercise, we suggest the reader to prove all these statements for $N=2$.
 The key relation follows from the condition $\p_u \Delta_3 (X) =0$:
\be
 \label{29}
 \p_v \biggl[ \p_v \biggl({X \over {\p_u X}}\biggr) \, / \,
 \p_v \biggl({{\p_u^3 X} \over {\p_u X}}\biggr) \biggr] = 0\ .
 \ee
 It follows that the expression in the square brackets is equal to
 an arbitrary function $A_0(u)$ and thus we have
 \be
 \label{30}
 \p_v \biggl[ \biggl({X \over {\p_u X}}\biggr) \, + \,
  A_0(u) \, \biggl({{\p_u^3 X} \over {\p_u X}}\biggr) \biggr] = 0\ .
 \ee
 Denoting the expression in the square bracket by $-A_1(u)$ and
 introducing the notation ${\cal U}_1(u) = A_1(u) / A_0(u)$ and
 ${\cal U}_0(u) = 1/A_1(u)$, we get the first of Eqs.(\ref{27})
 with $N=2$. Repeating similar derivation starting
 with $\p_v \Delta_3 (X) =0$ one can obtain the second of
 Eqs.(\ref{27}).

Let us return to the general solution of Eq.(\ref{21}).
 In fact, considering Eqs.(\ref{28}) as inhomogeneous differential
 equations for $a_{N+1}(u)$, $b_{N+1}(v)$ with arbitrary chosen functions
 $a_i(u)$, $b_i(v)$ ($1\leq i \leq  N$), it is easy to write the explicit
 solution of this problem:
 \be
 \label{31}
 a_{N+1}(u) = \sum_{i=1}^N a_i(u) \int^u_{u_0} d\bar u \ W^{-2}_N(\bar u) \
  M_{N,\,  i}(\bar u) \ .
 \ee
 Here $W_N \equiv W[a_1(u),..., a_N(u)]$  is the Wronskian of
 $N$ arbitrary chosen functions $a_i$ and
 $M_{N, \, i}$ are the complementary minors of the last row in the Wronskian.
 (Replacing $a$ by $b$ and $u$ by $v$ we can find the expression for $b_{N+1}(v)$
 from the same formula (\ref{31})). For the simplest $\cA_2$-case:
 \bdm
 a_3 (u) = \sum_{i=1}^2 a_i(u) \int^u_{u_0} {{d\bar u} \over {W^2_2(\bar u)}} \
  M_{2,\, i}(\bar u) \equiv  \int^u_{u_0} d\bar u \
  {{a_1(\bar u) a_2(u) - a_1(u) a_2(\bar u)} \over
    {(a_1(\bar u) a_2^{\prime}(\bar u) - a_1^{\prime}(\bar u) a_2(\bar u))^2}} \ .
 \edm
 Thus we have found the expression for the basic solution $X_1$ in terms
 of $2N$ arbitrary chiral functions $a_i(u)$ and $b_i(v)$.
 To complete constructing the solution we should
 derive the expressions for all $X_n$ in terms of $a_i$ and $b_i$.
 This can be done with simple combinatorics that allows one to express $X_n$
 in terms of the $n$-th order minors. For example, it is easy to
 derive the expressions for $X_2$:
  \be
  \label{31a}
 X_2 = \varepsilon_1 \Delta_2 (X_1) =
   \varepsilon_1 \sum_{i<j}  W[a_i(u), a_j(u)] \ W[b_i(v), b_j(v)] \ ,
 \ee
 which is valid for any $N \geq 1$ ($i,j = 1,...,N+1$).
 Note that expressions for all $X_n$
 have a similar separated form
 with higher-order determinants (see \cite{Leznov}).

 Our simple representation of the $\cA_N$ Toda solution is completely
 equivalent to
 what one can find in \cite{Leznov} but is more convenient for treating
 some problems. For example, it is useful in discussing asymptotic and
 analytic properties of the solutions of the original physical problems.
 It is especially appropriate for constructing wave-like solutions of
 the Toda system which are similar to the wave solutions of the
 $N$-Liouville model. In fact, quite like the Liouville model,
 the Toda equations
 support the wave-like solutions. To derive them let us first identify
 the moduli space of the Toda solutions. Recalling the $N$-Liouville
 case, we may try to identify the moduli space with the space of the
 potentials ${\cal U}_n(u)$, ${\cal V}_n(v)$. Possibly, this is not
 the best choice and, in fact, in the Liouville case we finally made
 a more useful
 choice suggested by the solution of the constraints.
 For our present purposes the choice of the potentials is as good as any
 other because each choice of ${\cal U}_n(u)$ and ${\cal V}_n(v)$ defines
 some solution and, vice versa, any solution given by the set of the
 functions ($a_1(u),..., a_{N+1} (u)$), ($b_1 (v),..., b_{N+1} (v)$)
 satisfying
 the Wronskian constraints (\ref{28}) defines the corresponding set of
 potentials (${\cal U}_0(u ),...,{\cal U}_{N-1}(u)$),
 (${\cal V}_0(v),...,{\cal V}_{N-1}(v)$).

 Now, as in the Liouville case, we may consider the reduction of the
 moduli space to the space of constant `vectors' $(U_0,...,U_{N-1})$,
 $(V_0,...,V_{N-1})$. The fundamental solutions of the equations (\ref{27})
 with these potentials are exponentials (in the nondegenerate case):
 $\exp (\mu_i u)$, $\exp (\nu_i v)$.
 Then $X_1$ can be written as (for simplicity we take $f_i >0$):
 \be
 \label{32a}
 X_1 = \sum_{i=1}^{N+1} a_i (u) b_i (v)
 = \sum_{i=1}^{N+1} f_i \, \exp (\mu_i u) \, \exp (\nu_i v)
 = \sum_{i=1}^{N+1} \exp (\mu_i u + u_i) \, \exp (\nu_i v + v_i) \ ,
 \ee
 where the parameters must satisfy the
 conditions (\ref{28}). Calculating the determinant $\Delta_{N+1}(X_1)$
 and denoting the standard Vandermonde  determinants by
 \bdm
 D_{\mu} \equiv \prod_{i>j} (\mu_i - \mu_j) \ , \qquad
 D_{\nu} \equiv \prod_{i>j} (\nu_i - \nu_j) \ ,
 \edm
 one can easily find that (\ref{28}) is satisfied if
 \be
 \label{32b}
 \sum_{i=1}^{N+1} \mu_i \, = \, \sum_{i=1}^{N+1} \nu_i = 0 \ ,
  \qquad  \prod_{i=1}^{N+1} f_i \ D_{\mu} \ D_{\nu} =
 \tilde{\varepsilon}_{N+1} \ .
  \ee
   By the way, instead of the last condition we could write the
 equivalent conditions (\ref{28}):
 \be
 \label{32c}
   \prod_{i=1}^{N+1} \exp u_i = w_a \ , \qquad
   \prod_{i=1}^{N+1} \exp v_i = w_b \ , \qquad
   w_a \cdot w_b
   = (D_{\mu} \ D_{\nu})^{-1} \tilde{\varepsilon}_{N+1}  \ ,
  \ee
 where $\exp u_i$ and $\exp v_i$  are not necessary positive
 (e.g., we can make $\exp u_i $ negative by supposing that $u_i$
 has the imaginary  part $i \pi$) but here we mostly consider positive
 $f_i$.

 In this reduced case we may regard the space of the parameters
 ($\mu_i$, $\nu_i$, $u_i$, $v_i$) as the new moduli space,
 in complete agreement with the Liouville case. Having the basic
 solution $X_1$ given by Eqs.(\ref{32a})-(\ref{32b}) it is not
 difficult to derive $X_n$ recursively by using (\ref{19}).
 For illustration, consider the simplest TL theory
 $\cA_1 \bigoplus \cA_2$.
 Then $X_2$ is given by (\ref{31a}) and (\ref{32a})-(\ref{32b}):
 \be
 \label{32d}
 X_2  = \varepsilon_2 (D_{\mu} \ D_{\nu})^{-1} \sum_{k=1}^3
 (\mu_i - \mu_j) (\nu_i - \nu_j) \
 \exp (-\mu_k u - u_k) \ \exp (-\nu_k v - v_k)  \ ,
 \ee
 where $(ijk)$ is a cyclic permutation of $(123)$
 (and thus $\mu_i + \mu_j = -\mu_k$,
     $\nu_i + \nu_j = -\nu_k$).
 The next step is to consider the constraints (\ref{f60}),
 where $Y_3$ is the solution of the Liouville equation
 with constant moduli.

  Now, using Eqs.(\ref{32a})-(\ref{32d})
  for the $\cA_1 \bigoplus \cA_2$ case,
   one can find that the
  constraints are equivalent to the following equations:
  \be
  \label{36}
  \sum_{i<j} (\mu_i -\mu_j) (\nu_j-\nu_j)[3\mu_k^2 - C_{\mu}] = 0 \ ,
   \qquad
  \sum_{i<j} (\mu_i -\mu_j) (\nu_j-\nu_j)[3\nu_k^2 - C_{\nu}] = 0 \ ,
  \ee
  \be
  \label{37}
  \mu_1^2 + \mu_2^2 + \mu_1 \mu_2  =  C_{\mu} \ , \qquad
  \nu_1^2 + \nu_2^2 + \nu_1 \nu_2  =  C_{\nu} \ ,
  \ee
  where the constants $C_{\mu}$ and $C_{\nu}$ represent the
  contribution of the Liouville
  term. Computing the sums in Eq.(\ref{36}) we find
  that Eqs.(\ref{36}) are equivalent to the relations
  \be
  \label{38}
  3[(\mu_1^2 + \mu_2^2 + \mu_1 \mu_2) - C_{\mu}] \sum \mu_i \nu_i = 0 \ ,
  \quad
  3[(\nu_1^2 + \nu_2^2 + \mu_1 \nu_2) - C_{\nu}] \sum \mu_i \nu_i = 0 \ ,
   \ee
  which are satisfied as soon as Eqs.(\ref{37}) are satisfied.

  It is not difficult to check that the potentials ${\cal U}_1(u)$,
  ${\cal V}_1(v)$ for the exponential solutions are
 \be
 \label{39}
 {\cal U}_1(u) = - (\mu_1^2 + \mu_2^2 + \mu_1 \mu_2)
 \equiv  \half \sum \mu_i^2 , \qquad
 {\cal V}_1(v) = - (\nu_1^2 + \nu_2^2 + \mu_1 \nu_2)
 \equiv \half \sum \nu_i^2 ,
 \ee
 and thus the constraints have an
 extremely simple and natural form:
 \be
 \label{40}
 {\cal U}_1 + C_{\mu} =0 \ , \qquad
 {\cal V}_1 + C_{\nu} = 0.
 \ee
 These constraints can easily be solved. In Appendix~2 we write
 the constraints for the general solutions in the
 $\cA_1 \bigoplus \cA_2$ theory in the same form but with
 ${\cal U}_1\,$, $C_{\mu}$ depending on $u$ and
 ${\cal V}_1\,$, $C_{\nu}$ depending on $v$.
 Unfortunately, this does not help to solve these more general
 equations.

\section{A simple integrable model of (1+1)-dimensional dilaton
gravity coupled to Toda scalar matter}

Here we consider a simple DG model (first briefly discussed in
\cite{ATF6}), in which the potential $V$ is independent of $\f$, i.e.,
 $\p_{\f}V \equiv 0$. Supposing, as above, that the potentials $Z_n$
 are constant (we take $Z_n \equiv -1$) it easy to see that the
 equations for the matter fields $\psi_n$ can be separated and
 solved independently
 from the other equations\footnote{
 In this section we slightly change notation and consider $N$ matter fields
 $\psi_n$, $n=1,...,N$.}. This obviously follows from the fact that
 the equation (\ref{F.18}) for the metric $f(u,v)$ defines the
 essentially
 trivial metric $f = \varepsilon \apr \bpr$, which can locally be transformed
 to $f = \varepsilon$. Thus, one has first to solve the matter equations
 (\ref{F.16}) and then the dilaton equation (\ref{F.15}), while the
 constraints (\ref{F.17}) give additional relations between the dilaton
 and matter fields.

 Let us define the potential $V$ as a multi - exponential function
 of $\psi_n$
 \be
 \label{e1}
 V\,=\, \varepsilon \sum_{n=1}^N\,2g_n\,\exp{\sum_{m=1}^N \,\psi_m\,a_{mn}}\,
 \ee
  Then the matter equations (equations of motion) are
 \be
 \label{e3}
 \d_u\,\d_v\, \psi_n\,=\, \varepsilon \,\sum_{m=1}^N\, \,a_{nm}\,g_m
 \exp{\sum_{k=1}^N\, \psi_k\, a_{km}}\,.\ee
 Now we can use for the description of the model the equations
 (\ref{11}) - (\ref{14}) if we set in them $\ep_n \equiv 1$.
 The equations for $A_{mn}$ are very simple to solve and we
 discuss them at the end of this Section. The matter equations
 (\ref{13}) are integrable in the Toda - Liouville case but they
 are also integrable in the pure Toda case, the simplest example
 being the $\cA_2$ Toda theory.

 When  Eqs.(\ref{13}) are integrable we can find the complete
 solution of the model, including the
 constraints. However,
 it is not necessary to suppose that Eqs.(\ref{13}) are integrable
 to find particular solutions of all the equations. So, suppose that
 we have found a solution of the system (\ref{13}) and show how to
 derive the dilaton field and solve the constraints.
 In fact, this is a very simple exercise.

 The general solution of Eq.(\ref{F.15}) can be written as
\be
 \label{33}
 \f = -\ve \int_0^u \int_0^v d\bu\, d\bv \, V[\psi (\bu , \bv)] +
 A(u) + B(v) ,
\ee
where $A(u)$, $B(v)$ are arbitrary functions. The constraints (\ref{F.17})
in this model have the form
\be
 \p_i^2 \f  = -\sum_{n=1}^N  (\p_i \psi_n)^2 , \,\,\,\,\,\,\, i=u,v \, .
 \label{34}
\ee
Using the equations of motion (\ref{F.16}) we easily derive
\be
 \p_i V  = \ve \p_j \sum_1^N (\p_i \psi_n)^2 ,
 \,\,\,\,\,\,\, (i,j)=(u,v) \,\, {\rm or} \,\, (v,u) \,,
 \label{35}
\ee
find $A(a)$, $B(b)$ in terms of $\psi$, and finally obtain:
\be
 \label{34a}
 \f = -\ve \int_0^u \int_0^v d\bu d\bv \ V[\psi (\bu , \bv)] -
 \int_0^u d\bar{\bu}  \int_0^{\bar{\bu}}  d\bu \ \Phi_u (\bu) -
 \int_0^v d\bar{\bv}  \int_0^{\bar{\bv}}  d\bv \ \Phi_v (\bv) +
  A^{\prime}(0) u + B^{\prime}(0) v \,,
\ee
 where we omitted the unimportant arbitrary term
 $A(0) + B(0) = \f(0,0)$ and denoted
 \bdm
 \Phi_u (u) \equiv \sum_1^N (\p_u \psi_n (u,0))^2 \,, \,\,\,\,\,\,\,\,\,\,\,
 \Phi_v (v) \equiv \sum_1^N (\p_v \psi_n (v,0))^2 \,.
 \edm
 This completes finding the solution of the complete system of this
 model {\bf provided that we know the general solution of
 Eqs.(\ref{F.16})}.

      Now, to get integrable equations for $\psi_n$ we take the potential
 (\ref{e1}) for which the equations (\ref{12}) - (\ref{14})
 (with $\epsilon_n \equiv 1$)
        can be reduced to integrable Toda equations.
 Therefore, we find the explicit analytical
 solution for the nontrivial class of dilaton gravity minimally
 coupled to scalar matter fields. This model is a very complex
 generalization of the well studied CGHS model (with free scalar
 fields and with the trivial potential $V=g$). Our model has
 much richer geometric properties and a
 very complex structure of
 the space of its solutions. In particular, the construction of the
 solutions with constant moduli of the previous section is fully applicable
 and directly gives the generalized wave-like solutions that include
 the one-dimensional reductions -- the static and cosmological solutions.

 The general $\cA_N$ case can be solved and studied using the results
 presented in this paper.
 The easiest case is $N=1$ (the Liouville equation for a  single $\psi$). The
 first simple but really interesting
 theory is the case of two scalar fields
 satisfying the $\cA_2$ Toda equations. Taking, for example,
 \bdm
 V = \exp{(\sqrt 3 \ \psi_1 - \psi_2)} + \exp{(2 \ \psi_2 )} ,
 \edm
 we find the simplest realization of the $\cA_2$ Toda dilaton gravity model
 whose complete solution can be obtained by the use of the above derivations.
 An interesting question is the following: is it possible to derive such models by
 dimensional reduction of some `realistic'  higher-dimensional theories?

\section{One-dimensional Toda -- Liouville systems}
 Here we consider the one-dimensional multi-exponential models and,
 especially, those that can be reduced to integrable Toda - Liouville
 theories. These models can be derived either directly from the
 higher dimensional (super)gravity theories\footnote{
 Various approaches to dimensional reductions and
 concrete low-dimensional models can be found in many papers published
 in the last twenty years, see, e.g.,  \cite{Hajicek}, \cite{Gibbons},
 \cite{CAF1}, \cite{NKS}, \cite{Lukas} - \cite{ATF3}, \cite{Lid},
 \cite{Venezia} and references therein.}
 or by further reducing
 the two-dimensional dilaton gravity.
 Having in mind the reduction of DG to dimension one
 we consider somewhat more general two-dimensional Lagrangians
 (we return to the
 notation of Section~2 for $\psi_n$)
 \be
  \label{a1}
 {\cal L}^{(2)}\,=\,\sqrt{-g}\,\biggl[ \,\f R(g)\,+\,V(\f,\psi)\, +\,
 \sum_{n=3}^{N_1}Z(\f)\, (\nabla \psi_n)^2\,+ \,
  \sum_{p=N_1+1}^N\,Z_p(\f,\psi)\, (\nabla \sigma_p)^2 \, \biggr]\,.
 \ee
 The difference between $\sigma_p$ and $\psi_n$ is that ${\cal L}^{(2)}$
 depends only on the derivatives of $\sigma_p$ but not on $\sigma_p$
 itself.
 The simplest reduction of ${\cal L}^{(2)}$  to the one-dimensional case
relevant for the description of black holes and of cosmologies can be set
into the following form (see, e.g. \cite{ATF2}), \cite{ATF5}):
 \be
   \label{a5}
{\cal L}^{(1)}=\,{1\over \bar l(\tau)}\, \biggl( \,-\dot\phi \dot F
+\sum_n {\dot \psi}_n^2-\sum_p\,\bar Z_p(\phi,\psi) \,{\dot \sigma_p}^2
\,\biggr) \, +\, \bar l(\tau)\,\ve\,e^F \, \bar V(\phi,\psi)\,.
 \ee
Here $\phi$ is related to the original dilaton $\f$ by the equation
$\phi'(\f)\equiv -Z(\f)$,
 $\bar l(\tau)$ is the new Lagrangian multiplier,
$\bar l(\tau)=l(\tau)\,\phi'(\f)$, and
\be
\bar V(\phi,\psi)={V(\f,\psi)\over \phi'(\f)}, \quad \bar
Z_p(\phi,\psi)\,=\,-Z_p(\phi,\psi)\,\phi'(\f).
\ee

Note that $\tau$ may be either the time, $t$, or the space, $r$,
coordinate and correspondingly $\ep$ is
 equal to $\pm1$, reflecting the
sign of the metric $f$ (see ref. [1] and, for a more detailed discussion,
refs. [3] and [4]).
  In what follows we interpret
Eq.(\ref{a5}) as a cosmological model and thus denote
 $\tau=t$. In fact, the cosmological Lagrangian of type (\ref{a5}))
 can be derived directly
from higher - dimensional supergravity theories
 (see, e.g. \cite{VDA1}, \cite{ATF1}, \cite{ATF2})).

 Now, one can see that  the equations of motion for the Lagrangian
 (\ref{a5}) can be solved w.r.t. the fields $\sigma_p(t)$.  Indeed,
 the canonical momenta $P_q$ corresponding to $\sigma_p(t)$
 are conserved:
 \be
 \label{a20}
{d\over dt}\,P_q\equiv {d\over dt} {\d {\cal L}^{(1)} \over
\d{\dot\sigma_q}} = -2 {d\over dt} (Z_q(\f,\psi) \,{\dot\sigma_q) }\,=\,
  {\d {\cal L}^{(1)} \over \d{\sigma_q}} \, \equiv \, 0,
  \quad N_1+1\leq q\leq N\,.
 \ee
 We also have ($\psi_1$ and $\psi_2$ are defined in terms of
 $\f$, $F$, as in Section~2):
\be
\label{a25}
P_i\equiv  {\d{\cal L}^{(1)}\over \d {\dot \psi_i}},\quad 1\leq i \leq
N_1,
\ee
and so we find the Hamiltonian corresponding to the Lagrangian (\ref{a5}),
\be
 \label{a30}
{\cal H}^{(1)}\,= \,{l(t)\over 4} \,\biggl\{\, \sum_{i=1}^{N_1} \ep_i
P_i^2 - \sum_{q=N_1+1}^N \, \bar Z_q^{-1}(\phi,\psi) \, P_q^2+4\ve \,e^F
\, \bar V(\phi,\psi)\,\biggr\},
 \ee
 where now the
 $P_q$ are the integration constants ($\ep_1 = -1$,
 $\ep_n = 1, \, 1 \leq n \leq N_1$ ).
 This means that the Lagrangian (\ref{a5}) can be replaced by the
 effective one,
 \be
 \label{a35}
{\cal L}^{(1)}\,=\,{1\over \bar l(t)}\, \sum_{n=1}^{N_1}\,{\dot
  \psi_n}^2 \,+\, \bar l(t)\, \ve \, V_{\rm eff}(\phi,
 \psi)
 \ee
 where $V_{\rm eff}$ is the effective potential,
 \be
 \label{a40}
V_{\rm eff}(\phi,\psi)\,=\,\ve \, e^F\, \bar V(\phi,\psi) -
\sum_{q=N_1+1}^N {P_q}^2 \bar Z_q(\phi,\psi) \,.
 \ee

 If this effective potential is of the exponential form (as in Section~2),
 we may apply our general approach developed for the treatment of the
two$-$dimensio\-nal systems. Thus from now on  we simply forget about the
 transformations above. The only difference is that
 the new potential $V$ is the sum of two terms
 (denoting $N_2 \equiv N-N_1$ we call this $(N_1, N_2)$-model):
 \be
 \label{a45}
 V=\sum_{i=1}^{N_1} \, 2g_i \exp \,q_i \,\,+
 \sum_{p=N_1+1}^N  2g_p\exp q_p^{(0)}
 \ee
 and thus $a_{1i}=1+a_i \,,\, a_{2i}=1-a_i \,,\, i= 1,...,N_1$,
 while $a_{1p} = a_p \,,\,
 a_{2p}=-a_p \,,\, p=N_1 +1 ,..., N$.

 When the equations of motion can be reduced to the Toda - Liouville
 system,   we can use the one-dimensional solutions of
 Section 4. To find the class of the DG theories for which this is
 possible, we should find $a_{mn}$ in terms of $A_{mn}$ and solve
 the constraint on $A_{mn}$ generalizing Eq.(\ref{b10})
 As the matrix $a_{mn}$ now has a different form,  the
 solution of the equations $\ha^T \hep \ha = \hA$,
 which are equivalent to the bilinear system
 \be
 \label{a4}
  -a_{1m}a_{1n} \,+\, a_{2m} a_{2n}
  \,+\,\sum_{k=3}^N \,a_{km}a_{kn}\,=\,A_{mn} \,,
\ee
  do not coincide with those of Section 4, although the derivations will
 be very similar.

 To illustrate the difference consider the simplest case where
 \bdm
 a_{mn}\,=\,\left ( \begin{array}{lll}
 1+a_1 & 1+a_2 & a_3 \\
 1-a_1&1-a_2&-a_3 \\
 a_{31}&a_{32} & a_{33}\\
 \end{array} \right ) \edm
 In this case, Eqs.(\ref{a4}) give six equations for six unknowns,
 $a_1, a_2, a_3$ and $a_{31}, a_{32}, a_{33}$. It is not difficult to
 find that the solution is given by (recall that $A_{mn} = A_{nm}$)
 \bdm
 4a_1=a_{31}^2-A_1\,, \qquad  4 a_2=a_{32}^2 -A_2\,,\qquad
 a_{33} = \sqrt A_3 \,,
 \edm
 \bdm
 (a_{32}-a_{31})\,=\,(A_{32}\,-\,A_{31})/\sqrt{A_3} \,, \qquad
 2a_3 = a_{31}a_{33} - A_{31} = a_{32}a_{33} - A_{32 }\,.
 \edm
 It is clear that one of the $a_{mn}$ is not defined by the equations
 (e.g., $A_{31}$) and that the two expressions for $a_3$ give the
 constraint on $a_{mn}$:
 \be
 \label{a2}
 (A_1\,+\,A_2\,-\,2A_{12}) A_3 \,=\,(A_{23}-A_{13})^2 \,.
 \ee
 This constraint is similar to Eq.(\ref{b10}) and has the same origin
 and meaning. It is not satisfied for the $\cA_3$ Cartan matrix but
 can be satisfied for the $\cA_1 \bigoplus \cA_2$ Cartan matrix
 (with $A_{21} = A_{31} =0$ and $A_1 = -3/2$).

 Let us now consider the general system (\ref{a4}), which
  is not much more difficult to study than the `standard' system
 (\ref{1}). For this reason we omit details and only give a brief
 summary of the
 main differences. Eqs.(\ref{1}), (\ref{1a})
 are valid for $m,n \leq N_1$ (we denote such $m,n$ by $i,j$).

 The new equations are (the vectors $V_m$ are defined as
 $V_m \equiv [a_{3m},...,a_{Nm}]$ for all $m$):
 \be
 \label{a7}
 -2a_p  + V_p\cdot V_i\,=\,A_{pi}\equiv A_{ip}\,, \qquad
 V_p \cdot V_q\,=\,A_{pq} \,, \qquad i\leq N_1 \,,\, N_1+1\leq p\leq N
 \ee
 The first system of equations (`$ip$-equations') is equivalent to
  ($k = 2,...,N_1$):
 \bdm
 V_p\cdot v_k \equiv V_p (V_k-V_1) = A_{ip} - A_{1p}\,, \quad
 -2a_p + V_p \cdot V_1 = A_{1p} \,.
 \edm
 Thus we get $N(N - 1)/2$ equations for $N-1$ vectors $v_k, V_p$
 in the  $(N-2)$-dimensional space:
\be
\label{a8}
v_k\cdot v_l \,=\,A_1+A_{kl}-A _{1k}-A_{1l}\,,\quad k,l=2,...,N_1\,,
\ee
\be
 \label{a9}
 v_k\cdot V_p \equiv A_{kp}-A_{1p}\,, \quad v_p\cdot V_q\,=\,A_{pq}\,,
 \quad k=2,..,N_1\,, \quad  p=N_1+1,...,N\,,
 \ee
 while the  other equations define $a_{m}$ in terms of these vectors and
 of the matrix elements $A_{mn}$:
 \be
 \label{}
 4a_i\,=\,V_i^2 - A_i\,,  \qquad   2a_p\,=\,\,V_p\cdot V_1 -  A_{p1}\,\,,
 \ee

 As we have $N-1$ vectors $v_k,\,V_p$ in the $(N-2)$-dimensional space,
 one of them is a linear combination of the other ones, e.g.,
\be
\label{a10}
v_2=\sum_{n=3}^{N_1} x_n v_n \,+\,\sum_{p=N_1+1}^N x_p V_p.
\ee
 It follows that the matrix elements $A_{mn}$ should satisfy one constraint
 that can be obtained by first deriving $x_n, x_p$ and then writing the
 expression for $v_2^2$ in terms of $A_{mn}$
 (see the quite similar derivation in Section~3)\footnote{If $N_1=2$ and $N_2$
 is arbitrary, the constraint can be easily
 solved because the only dependence on $A_1$ is contained in
 $v_2^2=A_1+A_2-2A_{12}$ (see also the example below).}.

To actually derive $a_{mn}$ in terms of $A_{mn}$ we should solve th
equations for $v_k,\,V_p$ in a manner described in Section~3,
 Eqs.(\ref{ex1})-(\ref{ex2})
 (solving the equations in a fixed coordinate system, then
 using rotational invariance for the unit vectors $\hv_k,\,\hV_p$,
 etc.). A simplest nontrivial example is given by the case $N_1=N_2=2$.
 We leave it as an exercise to the reader.

 Note in conclusion, that many examples of the one-dimensional Toda
 and Toda - Liouville equations related to the equations considered
 in this and in the previous sections were derived by reductions of higher
 dimensional theories for description of some black holes and cosmologies
  (see, e.g. \cite{Fre} - \cite{Sorin}, \cite{Lukas} - \cite{Pope},
  \cite{Lid}). We plan to discuss applications of our formal approach and
  results in a forthcoming paper.

 \section{Conclusion}
 Let us briefly summarize the main results and possible applications.
 We introduced a simple and compact formulation of the general
 (1+1)-dimensional dilaton gravity with multi-exponential potentials
 and derived the conditions allowing to find its explicit solutions
 in terms of the Toda theory. The simplest class of  theories
 satisfying these conditions is the Toda - Liouville theory.
 In Section~5 (see also \cite{ATF6}),
 we show that the models with the potentials
 independent of the dilaton $\f$ can be explicitly solved if $A_{mn}$
 is any Cartan matrix. In this case adding the Liouville part is
 unnecessary. We also proposed a simple approach to solving the
 equations in the case of the $\cA_N$ Toda part.

 Of special interest are simple exponential solutions derived in
 Section~4. They explicitly unify the static (black hole)
 solutions, cosmological models, and waves of the Toda matter coupled
 to gravity. Some of these solutions can be related to cosmologies
 with  spherical inhomogeneities or to evolving black holes but
 this requires special studies.
 Earlier we studied similar but simpler solutions in the
 $N$-Liouville theories in paper \cite{ATF5}. The main results of that
 paper, in particular the existence of nonsingular exponential solutions,
 are true also in the Toda - Liouville theory.

 Note that the
 one-dimensional Toda - Liouville cosmological models were
 met long time ago in dimensional reductions of higher-dimensional
 (super)gravity theories (see, e.g., \cite{Pope}). Considerations of the
 two-dimensional Toda - Liouville theories of this paper are equally
 applicable to
 the one-dimensional case. A preliminary discussion can be found in
 \cite{ATF6}, and the detailed consideration is presented
 in this paper,
 together with a detailed presentation of the results that were only
 briefly described there (some of the results of Sections~2-4 were
 recently presented in \cite{Quarks}).

 To include into consideration
 the waves one has to step up at least one dimension higher. The principal
 aim of the present paper was to make the first step and
 explore this problem in a simplest
 two-dimensional Toda environment.
 As a simple exercise (based on the results of this paper)
 one may consider the
  reductions from dimension (1+1) both to  dimension (1+0)
 (`cosmological' reduction) and to  dimension (0+1)
  (`static' or `black hole' reduction)
 as well as the moduli space reductions to waves. One of  the most interesting
 problems for future investigations is the connection between these three
 objects. It was discovered in the $N$-Liouville theory but now  we see
        that it can be found in a much more complex theory described by
       the Toda equations. It is not impossible that the connection also
       exists  (possibly, in a weaker form) in some nonintegrable theories.

 Finally, we must admit that the general problem of solving the
 energy an momentum constraints in the two-dimensional Toda - Liouville
 theory remains essentially unsolved. It is completely solved in the model of
 Sections~5 and in the one-dimensional model of Section~6 (as well
 as for the effectively one-dimensional solutions of Section~4).
 However, in the general two-dimensional Toda - Liouville models we cannot
 analytically satisfy the constraints for the general solution.
 To analytically solve this problem we should probably combine our analytic
 approach with the group-theoretical considerations of \cite{Leznov},
 \cite{Saveliev}.

 \section{Appendix}
 \subsection{Cartan matrices and $\alpha_{mn}$}
 For all Cartan matrices $A_{nn}=2$.
 For all Cartan matrices, except $\cG_2$ and $\cF_4$,
  \bdm
 A_{(n-1)n} = A_{n(n-1)} = -1, \qquad 2 \leq n \leq N-1 \,.
 \edm
 For the Cartan matrices of $\cA_N$, $\cB_N$, $\cC_N$,
 $\cD_N$, $\cE_N$
 (for the last series $N=6,7,8$):
 \bdm
 \cA_N , \cE_N : \,\,  A_{(N-1)N} = A_{N(N-1)} = -1 , \quad
 \cD_N : \,\, A_{(N-2)N} = A_{N(N-2)} = -1
  \edm
 \bdm
 \cB_N : \,\, A_{(N-1)N} = -2, \, A_{N(N-1)} =-1 ; \quad
 \cC_N : \,\, A_{(N-1)N} = -1, \, A_{N(N-1)} =-2 .
 \edm
 For $\cE_N$, in addition $A_{3N} = A_{N3} = -1$.
 The non-diagonal elements of $\cG_2$ are $A_{12} =-1$,
 $A_{21} = -3$. For $\cF_4$, all near-diagonal elements are
 equal to $-1$, except $A_{23} = -2$. We list only the
 nonvanishing elements, the  other ones are zero.

 The matrices of $\cA_N$, $\cD_N$, $\cE_N$ are symmetric
 and thus define the symmetric $\alpha_{mn}$.
 By $\tilde{\cB}_N$, $\tilde{\cC}_N$, $\tilde{\cG}_2$, $\tilde{\cF}_4$
 we denote  the symmetrized (as explained in the main text)
  Cartan matrices. For $\tilde{\cG}_2$ we have $A_{12} = A_{21} = -3$,
  $A_{11} = 2$,  $A_{22} = 6$. For $\tilde{\cF}_4$:
  \bdm
 A_{11} = A_{22} = 4 ,\,\,\,\,\,  A_{33} = A_{44} = 2 ,\,\,\,\,\,
 A_{12} = A_{23} = -2 ,\,\,\,\,\, A_{34} = -1 \,.
  \edm
 It is easy to derive
 $\alpha_{mn}$ using Eq.(\ref{16}). For  $\cA_N$: $\alpha_{nn} =-2$,
 $\alpha_{(n-1)n} = \alpha_{n(n-1)} =1$ ($2 \leq n \leq N-1$),
 and the  other elements vanish.
 For other series and for exceptional groups, the
 $\alpha_{mn}$ are
 not so trivial (e.g., $\alpha_{12}=1$, $\alpha_{21}=2$ for $\cC_2$) and
 thus our simple approach to solving the Toda equations
 is not directly applicable.

 \subsection{Constraint on general multi-exponential plus Liouville
 matrices $A_{mn}$}

 Let us consider symmetric
 $N \times N$ matrices $A_{mn}$ satisfying the conditions
 \be
 \label{01}
 A_{kl} = \delta_{kl} A_k \, , \quad 1 \leq l \leq L  \, , \quad
  1 \leq k \leq N \,.
 \ee
 For convenience, let us denote $A_{L+m, L+n} \equiv B_{mn}$ for
 $1 \leq m,n \leq  M \equiv N-L$, where $B_{mn}$ is an arbitrary
 symmetric $M \times M$ matrix. When $L \geq 2$ the constraint equation
 (\ref{b10}) for $A_{mn}$ has the following simple form:
 \be
 \label{02}
 (A_1+A_2)\,D\,=\,A_1 \sum_{p=3}^N\, D_p \,.
 \ee
 As the determinants $D_p$ are proportional to $A_1$ (because
 $(v_p \cdot v_q) \equiv A_1$) the r.h.s. is proportional to $A_1^2$.
 According to the theorem proven in Section~4, Eq.(\ref{02}) is linear
 in $A_1$ and thus the quadratic terms in both sides of the equation
 should cancel. It follows that it is sufficient to derive the determinant
 $D$. We can do this by induction in $L$.
 For $L=2,3,4$ it is easy to find that $D$ can be presented by the
 following simple formula:
 \be
 \label{03}
 D\equiv D(L)\,=\, \gamma_2 \biggl[\delta_M \bigl(\sum_{l=3}^L \gamma_l +
 \gamma_1 + \sigma_{M-1} \bigr) \biggr] \prod_1^L A_k \,,
 \ee
 where $\gamma_l \equiv A_l^{-1}$, $\delta_M \equiv \det \hB$, and
 $\sigma_{M-1}$ is the determinant of the $(M-1)\times(M-1)$ matrix
 $\hB^{\prime \prime}$ having the following matrix elements
 \bdm
 B^{\prime \prime}_{mn} = B_{mn} - B_{mM} - B_{Mn} + B_{MM} \,, \quad
 1 \leq m,n \leq M-1 \,.
 \edm
 Direct computation of $D(L+1)$ with the aid of Eq.(\ref{03}) now allows
 to prove that the expression (\ref{03}) is valid for any $L$.
 The last step is to insert $D$ into equation (\ref{02}) which gives:
 \be
 \label{04}
 (A_1+A_2)\,D\,=\, \biggl\{ \biggl[ \delta_M \bigl( \sum_{l=3}^L \gamma_l
 + \gamma_1 + \sigma_{M-1} \bigr) \biggr] + \delta_M \,\gamma_2 +
 A_1 \biggl( \gamma_2 \sum_{l=3}^L \gamma_l \,+\,
 \sigma_{M-1} \biggr) \biggr\} \prod_1^L A_k \,.
 \ee
 Neglecting the last term that is proportional to $A_1^2$, we finally find
 that Eq.(\ref{02}) gives the beautiful relation
 \be
 \label{05}
  \delta_M \sum_{l=1}^L \gamma_l + \sigma_{M-1}  = 0 \,,
  \ee
 which is our final result.

 Supposing (naturally) that $\delta_M \neq 0$ we see that
 $\sum_{l=1}^L \gamma_l = 0$ is possible if and only if
 $\sigma_{M-1} = 0$.
 Note that for $\cA_N$ Cartan matrices $\sigma_{M-1} \neq 0$.
 Most probably, this is also true for all Cartan matrices of
 the simple groups. For the generic matrices $\hB$ the condition
 $\sigma_{M-1} = 0$ can be solved and thus
 there exists a class of the matrices $\hB$ for which
 $\sum_{l=1}^L \gamma_l = 0$. This means that it is possible to
 solve the energy and momentum constraints for any given solution
 of the $B$-system, which unfortunately, is not integrable and
 cannot be solved analytically.

 \subsection{Energy and momentum constraints for
 $\cA_1 \bigoplus \cA_2$ theory}
 Here we show that the energy and momentum constraints can be
 written similarly to Eqs.(\ref{40}) also for the general solution
 of the $\cA_1 \bigoplus \cA_2$ theory:
  \be
 \label{06}
 {\cal U}_1(u) + C_{\mu}(u) = 0 \ , \qquad
 {\cal V}_1(v) + C_{\nu}(v) = 0 .
  \ee
 For our simplest example, the Toda solutions $X_1$ and
 $X_2$ are given by (\ref{f45}) and (\ref{31a}) with $N=2$:
  \be
 \label{07}
  X_1=\sum_{i=1}^3\,a_i(u) \,b_i(v) \,,
  \quad
  X_2 = \varepsilon_1 \Delta_2 (X_1) =
   \varepsilon_1 \sum_{k=1}^3 W_k \bar{W}_k \ ,
  \ee
 where $W_k \equiv W[a_i(u), a_j(u)]$, $\bar{W}_k \equiv W[b_i(v),
 b_j(v)]$ and $(ijk)$ is a cyclic permutation of $(123)$.
 The $C_u$ constraint (\ref{f60}) is equivalent to the relation
 \be
 \label{08}
 \tilde{C}_u \equiv  X_1^{''} X_2 + X_1 X_2^{''} -
 X_1^{'} X_2^{'} \,-\, 4   X_1 X_2 X_L^{''} / X_L \, = 0 \,,
 \ee
 where the prime denotes $\p_u$ and $X_L$ is any solution of the
 Liouville equation satisfying Eq.(\ref{23}), from which it follows
 that
 \be
 \label{09}
 X_L^{''} / X_L = a^{''}(u) / a(u)  \equiv \, \alpha (u) .
 \ee

 Using Eqs.(\ref{07}) - (\ref{09}) it is easy to find that
 \be
 \label{010}
 \tilde{C}_u = \sum_{m,k} b_m \bar{W}_k \, [a_m^{''} W_k -
 a_m^{'} W_k^{'} + a_m W_k^{''} -4 \alpha W_k ] \,.
 \ee
 Let us first derive the `diagonal' part of the sum, using the
 definition of $W_k$ and Eq.(\ref{28}) for $N=2$:
 \be
 \label{011}
 \sum_{m=k} [...] = \sum_k b_k \bar{W}_k \,
 [(a_i a_j^{'''} - a_i^{'''}  a_j ) \, a_k  +
  (a_i^{''} W_{jk} + a_j^{''} W_{ki} + a_k^{''} W_{ij})
  - 4 \alpha W_k \,] \,.
 \ee
 Recalling the definition of ${\cal U}_1(u)$ (see (\ref{28a}))
 we find that the first term in the square brackets is equal
 to $-{\cal U}_1(u) a_k W_k$.  The second term in the square brackets
 is equal to the constant $w_a$, which is the Wronskian of $a_i (u)$
 (see (\ref{28})). The sum $w_a \sum_k b_k \bar{W}_k$ identically
 vanishes because
 \bdm
 \sum_k b_k \bar{W}_k = (b_1 b_2^{'} - b_1^{'} b_2) b_3 +
 (b_2 b_3^{'} - b_2^{'} b_3) b_1 +
 (b_3 b_1^{'} - b_3^{'} b_1) b_2 \equiv 0 \,,
 \edm
 and thus the sum (\ref{011}) is equal to
 \be
 \label{012}
  \sum_{m=k} [...] = -({\cal U}_1(u) + 4 \alpha (u))
 \sum_k a_k b_k W_k \bar{W}_k \,.
  \ee
 The `non-diagonal' part of the sum (\ref{010}) can be transformed to
 \be
 \label{013}
  \sum_{m \neq k} [...] = -({\cal U}_1(u) + 4 \alpha (u))
 \sum_{m \neq k} a_m b_k W_m \bar{W}_k \,
  \ee
 by employing for $m \neq k$ the evident identities
 \be
 \label{014}
 a_m^{''} W_k - a_m^{'} W_k^{'} + a_m W_k^{''} =
 - {\cal U}_1(u) \, a_m (u) \, W_k \,,
   \ee
 which can easily be checked.
 Collecting the diagonal and non-diagonal
 parts and returning to $X_1$ and $X_2$, we find that the
 $\bar{C}_u$ constraint (\ref{08}) is equivalent to
 \be
 \label{015}
 ({\cal U}_1(u) + 4 \alpha (u)) X_1 X_2 = 0 \,.
   \ee
 Repeating the same derivation for $C_v$ and recalling that
 $X_1 X_2 \neq 0$, we complete the proof of Eqs.(\ref{06}),
 with the evident notation $C_{\mu} (u) \equiv 4 \alpha (u)$,
 $C_{\nu} (u) \equiv 4 \beta (v)$.

  Note that in our derivation the number of the
  Liouville components may be arbitrary. On the other hand, we
 essentially used the simple structure of the $\cA_2$ solutions
 and thus our derivation cannot be directly applied to other
 the higher $\cA_N$ equations.


\bigskip
\bigskip

 {\bf Acknowledgment:}

 One of the authors appreciates financial support from
 the Department of Theoretical Physics of the University of Turin and INFN
(Turin Section), where this work was completed.

 This work was supported in part by the Russian Foundation for Basic
 Research (Grant No. 06-01-00627-a).


\end{document}